# *Enabling comparison of UrQMD with Geant4 hadronic models*

**Khaled Abdel-Waged[1], Nuha Felemban[1,2] and V.V. Uzhinskii[3]**

[1]*Umm Al-Qura University- Faculty of Science-Physics Department. E. mail: khelwagd@yahoo.com*
[2]*King Saud University- Faculty of Science-Physics and Astronomy Department.*
[3]*The European organization for nuclear research (CERN).*

# Abstract

Geant4 has an abundant set of physics models that handle the diverse interaction of particles with matter across a wide energy range. However, there are also many well established reaction codes currently used in the same fields where Geant4 is applied. One such code is the ultra-relativistic quantum molecular dynamic (UrQMD) model. In order to take advantage of the UrQMD code, we create a tool to enable comparisons among UrQMD and Geant4 hadronic models. This tool allows a user to process the output file of UrQMD through Geant4 toolkit, while at the same time, can choose among different Geant4 hadronic model generators. As an example, the UrQMD model is compared with the HARP-CDP experimental data and with the Binary and Fritiof generators, in the framework of Geant4. It is shown that the UrQMD model can better reproduce charged pion production for p+Cu and Pb interactions at 3, 8 and 15 GeV/c, and is a good candidate for Geant4 hadronic models.

# 1.Introduction

The increasing size and complexity of the software of high energy physics experiment requires the development of object oriented programming written in C++. Geant package is one such example [1]. The Geant4 simulation toolkit can handle the physical layout of an experiment; including tracking, geometry description, detector responses and digitalization of the experiment as well as management of events processing. The physics (hadronic) modeling of Geant4 cover diverse interactions over an extended energy range from thermal neutrons to Large Hadron Collider (LHC) experiments.

Recently, results of the Geant4 hadronic models have been compared with the HARP-CDP experimental data of secondary protons and charged pions (at production angles $20 < \Theta < 125^0$), in the interactions with targets spanning the full periodic table of elements, of proton and pion beams with momentum from 3 to 15 GeV/c [2]. Two types of intra-nuclear cascade (INC) models have been used: the Bertini model and the Binary model (up to ~10 GeV). At higher energies, two parton string models, the quark gluon string (QGS) and the Fritiof (FTF) models, are applied. The comparison of HARP-CDP data and the Geant4 hadronic models has shown that none of the models can fully describe the data.

To overcome this problem, we create a tool to enable processing of the results of ultra-relativistic quantum molecular dynamis (UrQMD) model [3] in Geant4. The performance of the UrQMD model is examined by comparing with the HARP-CDP experimental data and with Geant4 hadronic models. The main aim is to show that the UrQMD model is an appropriate candidate for the Geant4 hadronic models.

In Sec.2, a brief description of the Binary, Fritiof and UrQMD codes are provided. Sec.3 gives an explanation of how the results of the UrQMD code are processed in Geant4 package. A comparison of secondaries produced from UrQMD, Binary and Fritiof generators, in the framework of Geant4, with HARP-CDP experimental data is presented and discussed in Sec.4. Finally, we summarize our findings in Sec.5.

## 2.Reaction codes

### 2.1 UrQMD model

The UrQMD model was developed in 1998 with the intent to provide a unified description of various aspects of nuclear reactions. It is designed to cover the best possibilities of microscopic transport theoretical calculations in the energy range between 100 MeV/A and 200 GeV/A. At (1-10 GeV) energies, all baryonic resonances up to an invariant mass of 2.25 GeV as well as mesonic resonances up to 1.95 GeV, as tabulated by the Particle Data Group, are taken into account. In order to deal with the excitation and decay of color strings, above the resonance region, the (original) Fritiof model [4] is incorporated into the UrQMD code. All of the hadronic states can propagate in straight trajectories and re-interact in phase space.

A detailed description of the UrQMD is available in Ref.[3] and the source code of UrQMD can be download from http://urqmd.org/.

**2.2 Binary model**

The Binary model [5], abbreviated under the name "Binary" in Geant4, describes the interactions of protons and neutrons with nuclei as binary collisions between a primary or secondary particle and an individual nucleon of the nucleus. Unlike UrQMD model, collisions between participants are not considered. Particles entering the nucleus are propagated in a time-independent nuclear potential. Outside the nucleus, particles travel along straight trajectories. In analogy with the UrQMD model: (a) Experimental data and PDG parameterizations are used in the calculations of proton-proton (pp) and proton-neutron (pn) total, inelastic and elastic cross sections; (b) Delta resonances with masses up to 1.95 GeV and excited nucleons with masses up to 2.25 GeV are taken into account; (c) the partial and total decay width of a resonance is taken as mass dependent; (d) the elastic medium modified angular distributions of the scattered particles, other than nucleon-nucleon scattering, are calculated from the collision term of relativistic Boltzmann Uehling Uhlenbeck (RBUU) equation [6].

## 2.3 Fritiof model

The Fritiof model is implemented in Geant4 under the abbreviation FTF. The original Fritiof model assumes that in hadron-nucleus (hA) and nucleus-nucleus (AA) collisions, nucleons are excited in primary collisions and can interact with each other and with other nucleons, thus increasing their mass. Excited hadrons are considered as QCD strings, whose fragmentation produces hadrons. A significant reversion of the model is made by Geant4 hadronic group in order to tune the model parameters for description of hA and AA interactions [7]. Improvements in the dynamical contents of the model include: (a) cascading of secondary interactions (in the framework of Reggeon theory inspired model); (b) a separate simulation of diffractive and non-diffractive collisions, and (c) a treatment of the secondary particle formation time. These modifications allow to use FTF down to about 3-5 GeV/c.

## 3. Enabling the results of the UrQMD code to hadronic framework of Geant4

There are three basic types of modeling in Geant4, data driven, parameterization driven and theory driven modeling. Their possible realizations in the object oriented component system of Geant4 follow the Russian dolls approach to implementation framework design. In the following, we describe the usage of the data driven approach in modeling the results of the UrQMD code in Geant4.

First of all, a user will be asked to obtain an output file from the UrQMD. Of all possible output files of the UrQMD, the output file "file19" is chosen for processing secondaries of the interactions. The file19 output complies with the open standard codes and routines (OSCAR) format.

Two sub-directories are created for processing file19 in Geant4 toolkit, namely "UrQMD" and "Mytest". The contents of these directories are shown in Fig.1.

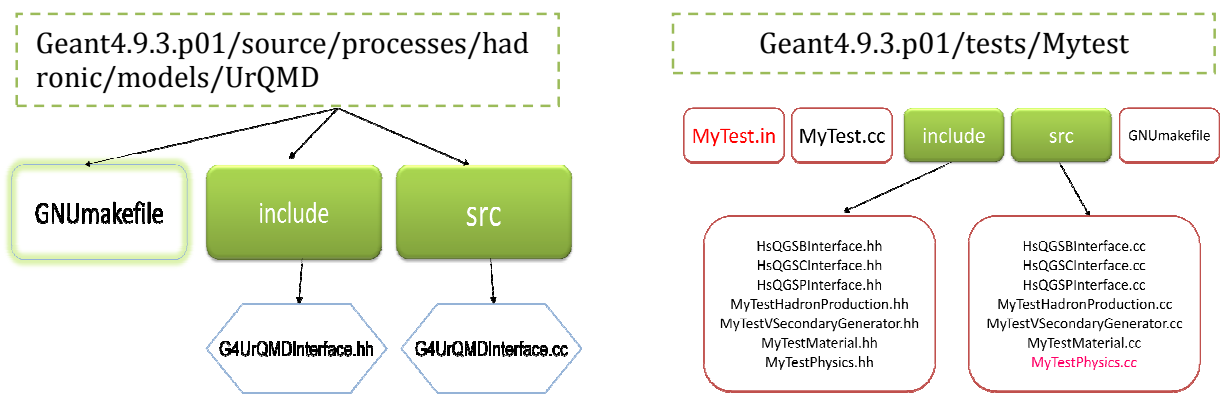

Fig.1. A schematic view of the contents of the sub-directories that a user should create for processing the output of the UrQMD through Geant4 toolkit.

In order to run the code one needs to define the running parameters with an input file. The input file "Mytest.in" make "UrQMD" generator accessible to a user. The generator is called via MyTestPhysics.cc file.

In G4UrQMDInterface.cc file, the final state particles of UrQMD code are processed in the ApplyYourself() method of the corresponding model class. The cross sections, final state production and isotope production are handled by the G4HadronicProcess class.

The following lines are introduced in Gnumakefile:

*name:=G4hadronic_UrQMD*

*-I$(G4BASE)/processes/hadronic/models/UrQMD/include \*

The line:

*-I$(G4BASE)/processes/hadronic/models/UrQMD/include \*

is also written in binmake.gmk file which is located at

*/geant4.9.3.p01/config/*.

It should be pointed out that when installing Geant4 toolkit a user is advised to enable both the "compound" and "granular" libraries. The line: *export G4Lib_USE_Granular=1* should be exported in the *.bash_profile* file of the system.

The code was successfully compiled and run. The contents of "UrQMD" and "Mytest" directories are available on request.

## 4. Numerical test and results

In this section, we compare the charged pion spectra predicted by UrQMD, Binary and FTF models, in the framework of Geant4, for p+$^{64}$Cu and $^{208}$Pb interactions at 3, 8 and 15 GeV/c. In the numerical calculations, the default UrQMD (version 1.3) is running in the cascade mode and no adjustments have been attempted.

The present investigation was carried out on the following platform:

*OS: Red Hat Linux 4.1.2-64*

*Compiler: gcc 4.1.2*

*Geant4: ver.9.3.p01.*

*Processor:Intel® Pentium® M Processor 1700 MHZ.*

We performed 50000 simulations at various impact parameter from 0 to R+0.5 fm, where R is the target radius.

| p+Cu | $CPU_{FTF}$ | $CPU_{Binary}$ | $CPU_{UrQMD}$ |
|---|---|---|---|
| 3 GeV/c) | 0m55s | 17m55s | 38m03s |
| 8 GeV/c) | 1m19s | 18m02s | 41m01s |
| 15(GeV/c) | 5m7s | 20m05s | 44m12s |

| p+Pb | $CPU_{FTF}$ | $CPU_{Binary}$ | $CPU_{UrQMD}$ |
|---|---|---|---|
| 3 GeV/c) | 2m52s | 77m59s | 69m18s |
| 8 GeV/c) | 2m59s | 88m52s | 93m14s |
| 15(GeV/c) | 6m3s | 90m05s | 103m31s |

**Table1.** The CPU-time of the running generators for 50000 simulations.

The CPU-time of the model generators are shown in Table 1 for the studied reactions. As one can see, the UrQMD model takes relatively longer CPU-time compared to Binary model. In contrast, FTF is ~18 times less CPU-time.

In Figs.2-5, we test the applicability of UrQMD model to charged pion production from p+Cu and Pb interactions at 3, 8 and 15 GeV/c. The solid histograms denote the UrQMD calculations. In the same figures we show the predictions of Binary (dashed histograms) and FTF (solid lines) models. As one can see, significant variation of the charged pion production with the Binary and FTF calculations is found. In particular, the Binary calculations tend to be higher than the experimental data in the region of 0.1-0.3 GeV at incident momentum of 3 GeV/c for the specified reactions. This overestimation becomes pronounced as the mass number of the target increases. On the other hand, as the incident momentum increases, the Binary calculations underestimate the charged pion spectra at $\Theta \leq$ 60-75$^0$. As for FTF, the charged pion spectra are overestimated in the region of 0.1-0.3 GeV when both the mass number and incident momentum increases. At high transverse momentum (>0.5 GeV/c), the pion spectra are underestimated as both the mass number and incident momentum decrease. In contrast, the UrQMD calculations show an overall good agreement with the experimental data, while the results underestimate the data at $\Theta \leq$ 30-40$^0$. It is known that the final multiplicity of pions depends largely on the description of not only its production but also its re-absorption during the transport process. The pion re-absorption is dominated by the re-absorption of the $\Delta$-resonance through $N + \Delta \rightarrow N + N$, which is obtained by the detailed balance from the cross section $\sigma_{NN \rightarrow N\Delta}$. Thus, the small pion spectra at forward angles ($\Theta \leq 30 - 40^0$) result from the higher $\sigma_{N\Delta \to NN}$ cross sections in UrQMD model. Indeed it is shown in Ref. [8] that the UrQMD model gives higher $\sigma_{N\Delta \to NN}$ cross section as a function of N$\Delta$ centre of mass energy compared to the experimental data and other transport models.

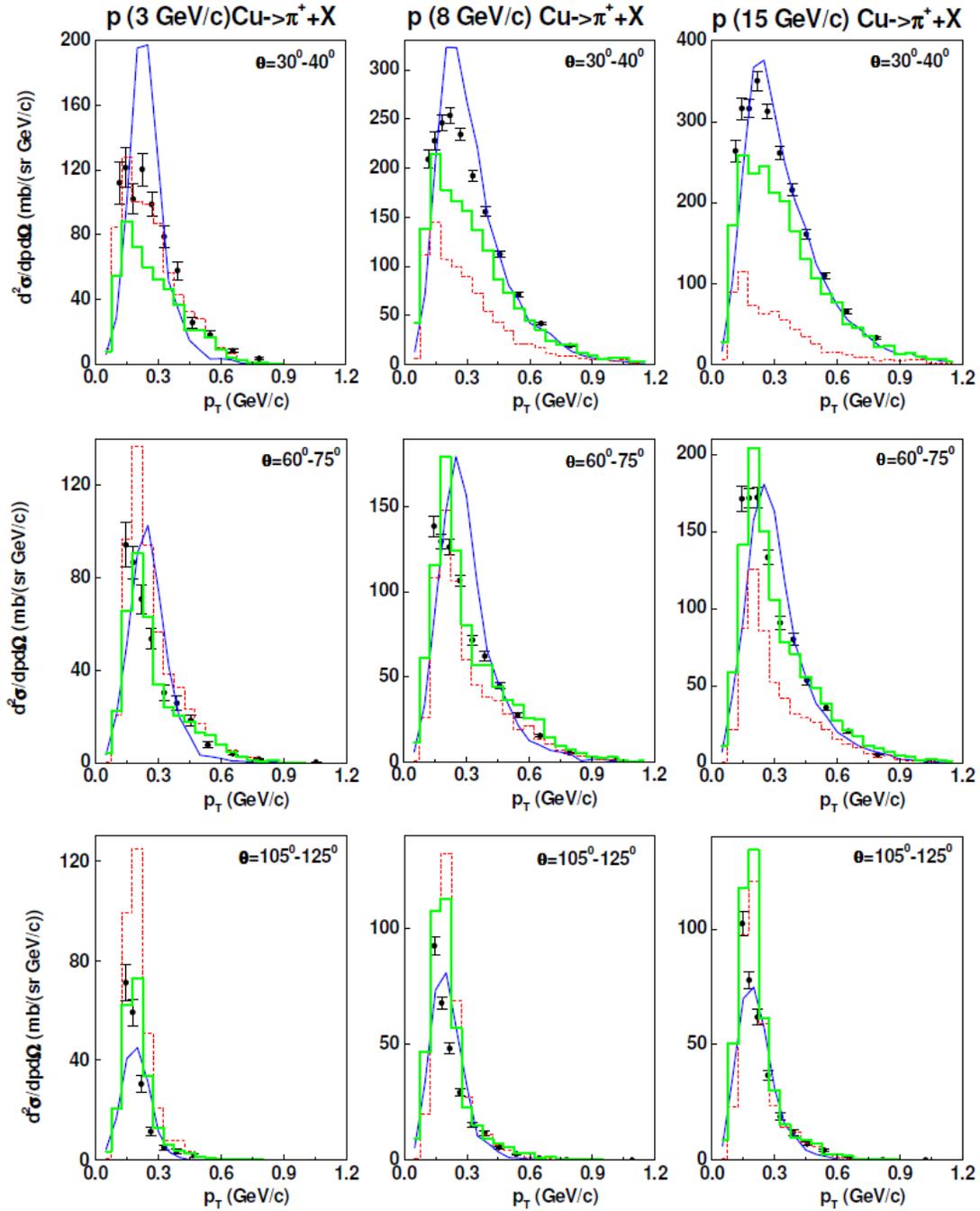

Fig.2 Inclusive cross sections for $\pi^+$ production from proton particles bombarding a copper target as a function of transverse momenta $p_T$ of secondaries. Solid points are the HARP-CDP experimental data and lines are the Geant4 calculations using different model generators (see text).

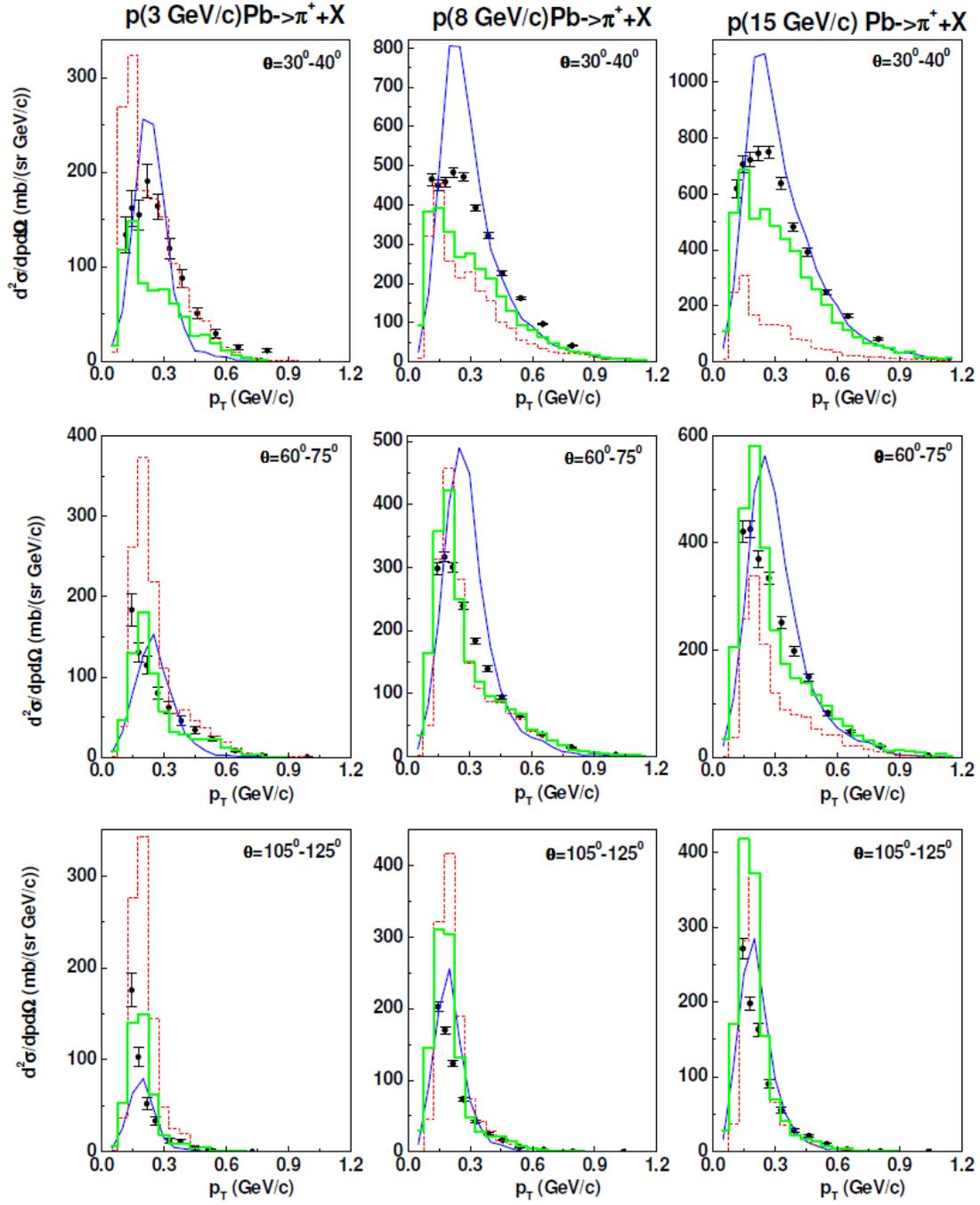

Fig.3. The same as Fig.2 but for p+$^{208}$Pb interactions.

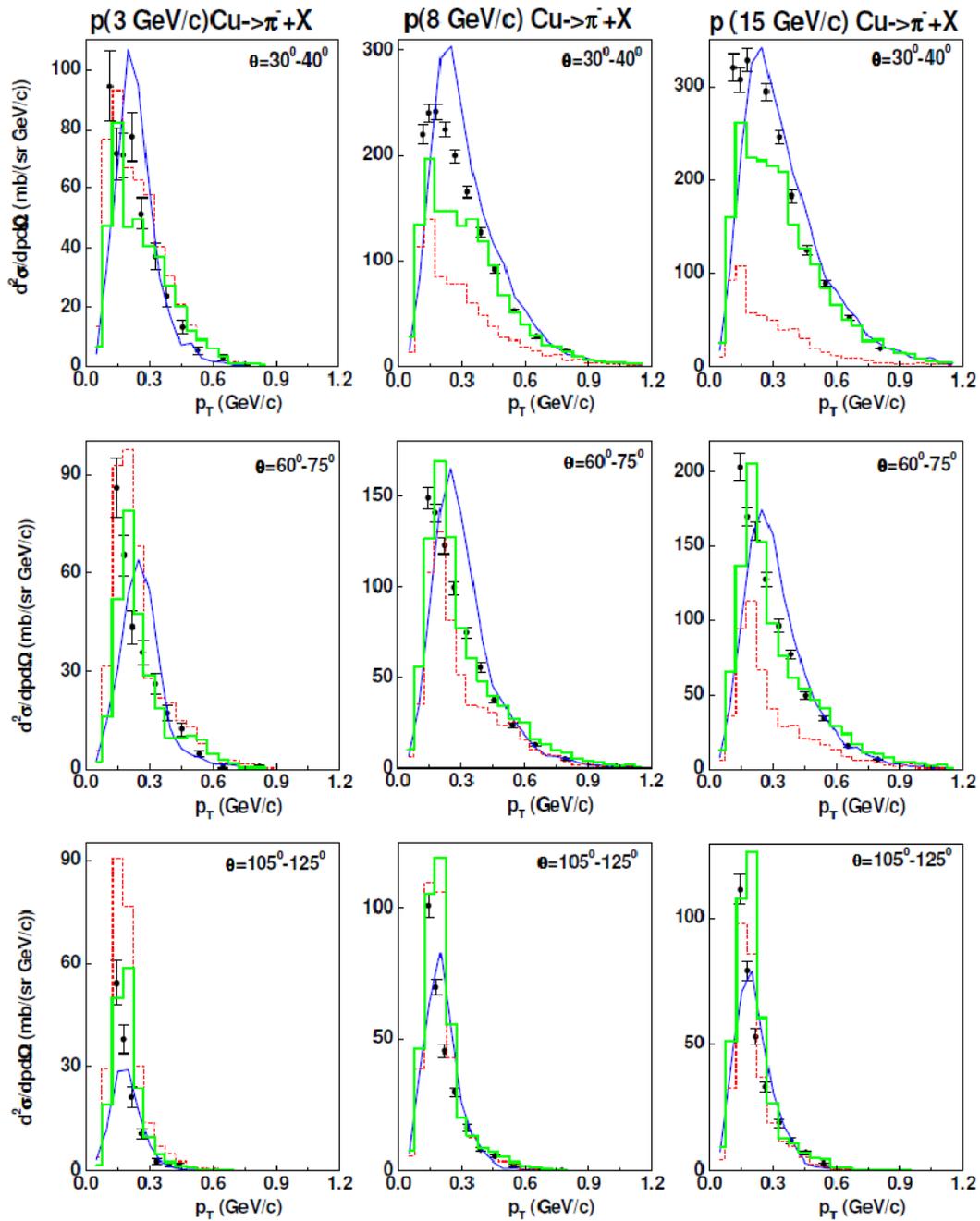

Fig.4. The same as Fig.2 but for $\pi^-$ production in p+$^{64}$Cu interactions.

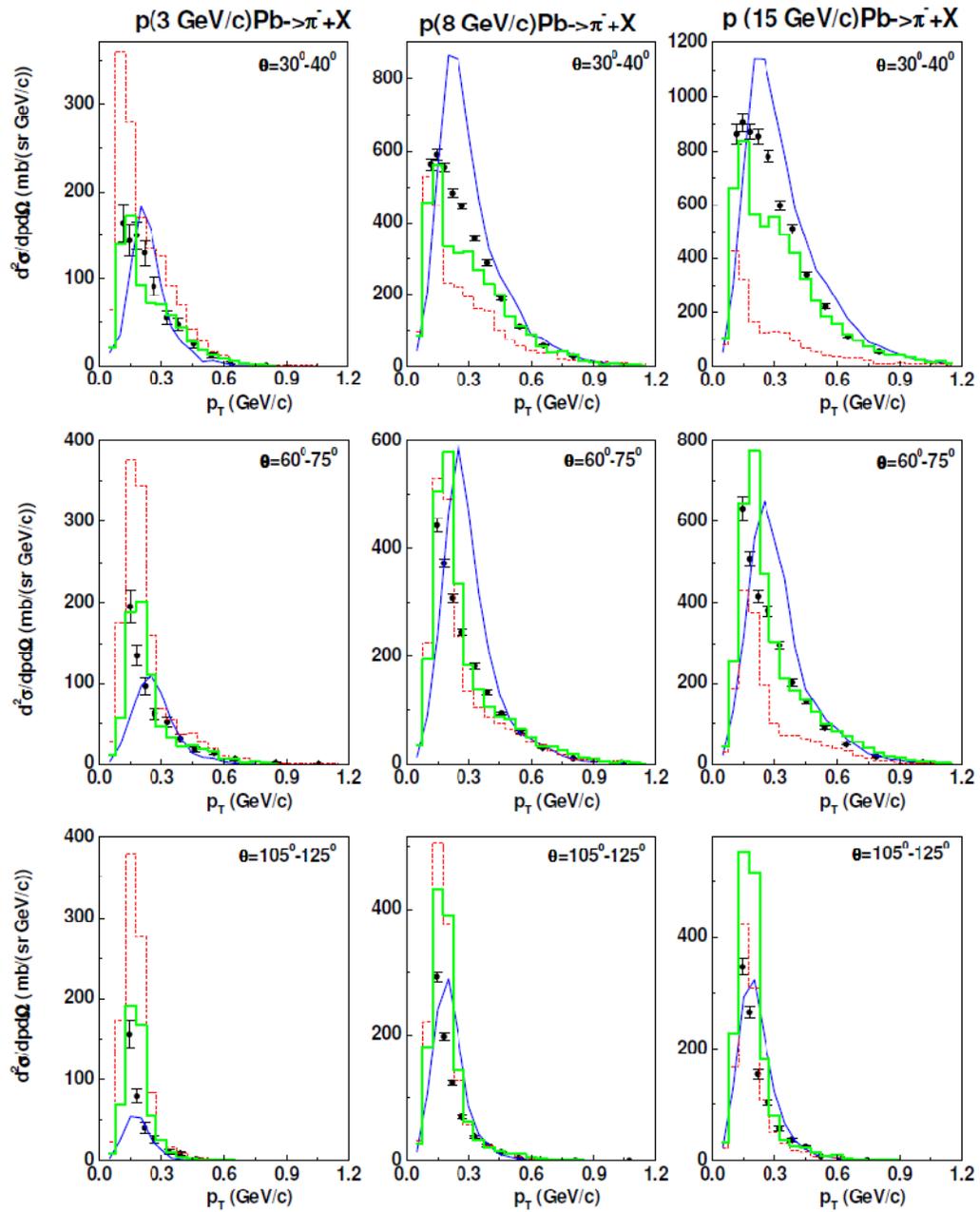

Fig.5. The same as Fig.2 but for $\pi^-$ production in p+$^{208}$Pb interactions.

## 5. Conclusions

In the present report, we enable a comparison between the UrQMD model and Geant4 hadronic models. This is done by allowing a user to process the output file (file19 (secondaries of interactions)) of UrQMD through Geant4 toolkit. The code has been validated against HARP-CDP experimental data and the Geant4 hadronic models for charged pion production in p+Cu and Pb interactions at incident beam momentum 3 GeV/c to 15 GeV/c. Even though the UrQMD calculation results are better than Geant4 hadronic models in the specified reactions, further improvements of the UrQMD model are needed, especially in the forward direction ( at $\Theta \leq 30\text{-}40^0$). Moreover, it would be desirable to build a full interface between the UrQMD model and Geant4 toolkit. This work is in progress.

# Acknowledments

This work is supported by the King Abdul-Aziz City for Science and Technology, the National Centre of Mathematics and Physics, Saudi Arabia, contract number 31-465. Kh Abdel-Waged and N. Felemban are thankful to the members of Geant4 hadronic group for stimulating discussions and help.